\documentclass[sigconf]{acmart}

\usepackage{dirtytalk}
\usepackage{enumitem}
\usepackage{multirow}
\usepackage{listings}
\usepackage{xcolor}
\usepackage{makecell}
\usepackage{array}
\usepackage{colortbl}

\usepackage{tcolorbox}
\usepackage{listings}
\definecolor{cardbackground}{RGB}{248,248,248}
\definecolor{cardtitle}{RGB}{33,33,33}
\definecolor{codebackground}{RGB}{245,245,245}
\definecolor{codekeyword}{RGB}{0,0,255}
\definecolor{codestring}{RGB}{163,21,21}
\definecolor{codecomment}{RGB}{0,128,0}
\newtcolorbox{feedbackcard}[1]{
    colback=cardbackground,
    colframe=gray!50,
    fonttitle=\bfseries\normalsize,  
    title=#1,
    coltitle=cardtitle,
    boxrule=0.5pt,
    arc=3mm,
    left=2pt,
    right=2pt,
    top=2pt,
    bottom=2pt,
    before skip=5pt,
    after skip=5pt,
    before upper=\footnotesize,
}

\AtBeginDocument{%
  }
\setcopyright{acmlicensed}
\copyrightyear{2018}
\acmYear{2018}
\acmDOI{XXXXXXX.XXXXXXX}

\acmConference[Koli Calling '25]{25th International Conference on Computing Education Research}{November 11--16, 2025}{Koli, Finland}
\acmISBN{978-1-4503-XXXX-X/2018/06}

\begin{document}
\title[Student Engagement with GenAI's Tutoring Feedback]{Student Engagement with GenAI's Tutoring Feedback:\\A Mixed Methods Study}

\author{Sven Jacobs}
\orcid{0009-0000-5079-7941}
\affiliation{
  \institution{University of Siegen}
  \city{Siegen}
  \country{Germany}
}
\email{sven.jacobs@uni-siegen.de}

\author{Jan Haas}
\orcid{0009-0006-3763-5976}
\affiliation{
  \institution{University of Siegen}
  \city{Siegen}
  \country{Germany}
}
\email{jan3.haas@student.uni-siegen.de}

\author{Natalie Kiesler}
\orcid{0000-0002-6843-2729}
\affiliation{
   \institution{Nuremberg Tech}
   \city{Nuremberg}
   \country{Germany}
}
\email{natalie.kiesler@th-nuernberg.de}

\renewcommand{\shortauthors}{Jacobs, Haas, Kiesler}

\begin{abstract}
How students utilize immediate tutoring feedback in programming education depends on various factors. Among them are the feedback quality, but also students' engagement, i.e., their perception, interpretation, and use of feedback. However, there is limited research on how students engage with various types of tutoring feedback.
For this reason, we developed a learning environment that provides students with Python programming tasks and various types of immediate, AI-generated tutoring feedback. The feedback is displayed within four components.
Using a mixed-methods approach (think-aloud study and eye-tracking), we conducted a study with 20 undergraduate students enrolled in an introductory programming course. Our research aims to: (1) identify what students think when they engage with the tutoring feedback components, and (2) explore the relations between the tutoring feedback components, students' visual attention, verbalized thoughts, and their immediate actions as part of the problem-solving process. 
The analysis of students' thoughts while engaging with 380 feedback components revealed four main themes: students express understanding or disagreement, additional information needed, and students explicitly judge the feedback. 
Exploring the relations between feedback, students' attention, thoughts, and actions showed a clear relationship. While expressions of understanding were associated with improvements, expressions of disagreement or need for additional information prompted students to collect another feedback component rather than act on the current information.
These insights into students' engagement and decision-making processes contribute to an increased understanding of tutoring feedback and how students engage with it. Thereby, this work has implications for tool developers and educators facilitating feedback. 
\end{abstract}

\ccsdesc[500]{Social and professional topics~Computer science education}

\keywords{Programming Education, Feedback, Large Language Models, Generative AI, GenAI}

\maketitle

\section{Introduction}
In pedagogical psychology and research, feedback is considered one of the most powerful factors for learning~\cite{hattietimperley2007,shute2008focus}.
According to Narciss~\cite{narciss.2006,narciss.2008}, several conditions and factors influence feedback processing and effects on learners. Among them are the feedback contents/types, their scope/function, presentation, and context. In addition, the feedback recipient plays a crucial role. Individuals process feedback differently from each other, or than expected by educators and tool developers. This is particularly relevant in computing education in Germany, where students are increasingly heterogeneous in terms of their prior education, competencies, programming experience, and feedback needs~\cite{kiesler2024modeling}.

With the emergence of GenAI and related tools, their potential for providing individual feedback at scale has become subject to research ~\cite{becker.2023a,kiesler.2023c,hellas.2023,phung2023generativeaifor}. Qualitative analyses have shown the increasing completeness and accuracy of recent GenAI tools' feedback for programming tasks and student solutions~\cite{azaiz.2024, koutcheme.2024a, lohr.2025a}. There is no doubt that novices are using GenAI tools in the context of introductory programming courses. Recent research has, for example, focused on how students chat with GenAI tools by analyzing chat protocol and students' perspectives~\cite{scholl2024analyzing,scholl2024noviceprogrammersuseexperience,scholl2025students}. Yet, a scoping literature review noted a research gap regarding students' perceptions and interactions with GenAI feedback in introductory programming education~\cite{stone_exploring_2024}. Moreover, it is crucial to consider if and how students acknowledge and process GenAI feedback before concluding on its usefulness or effects on learners. 

To address this gap, we investigate programming students' engagement with GenAI feedback types as part of an online learning environment for Python exercises. We analyze students' attention, perceptions, and immediate actions in response to GenAI feedback via a mixed-methods approach including eye-tracking and a think-aloud protocol. The present work is guided by the following research questions: 

\begin{enumerate}
    \item[\textbf{RQ1}]
    \textit{What do learners think aloud when they engage with AI-generated tutoring feedback components in an online learning environment for Python programming tasks?}
    \item[\textbf{RQ2}]
    \textit{What are the relations between tutoring feedback components, and students' (visual) attention, (verbalized) thoughts, (immediate) actions?}
\end{enumerate}

The \textbf{goal} of this empirical research study is to gain a qualitative understanding of how students engage with different types of AI-generated GenAI, i.e., how they acknowledge and process it, and, eventually, act upon it. Understanding students' perspectives is crucial for educators' pedagogical practices, and the integration of GenAI tools for feedback at scale. It further has implications for tool developers interested in improving their instructions, user interface, and other aspects of human-computer interaction.  

\section{Background and Related Work}

\citet{shute2008focus} defines formative feedback as \say{information communicated to the learner that is intended to modify the learner’s thinking or behavior for the purpose of improving learning}. The primary goal is not to grade or evaluate, but to support the ongoing learning process. 
Elaborated tutoring feedback (specificity of formative feedback) provides information about particular responses or behaviors beyond their accuracy and usually addresses the correct answer~\cite{shute2008focus}.
The goal is to guide the learner towards successful task completion ~\cite{narciss.2008}. In the following, we will outline the implementation of feedback in learning environments before and after the GenAI revolution. We further provide insights into recent research on students' engagement with feedback. 

\subsection{Implementation of Feedback in Learning Environments for Programming Education}

\citet{keuning.2019} systematically reviewed and analyzed 101 tools with programming exercises to identify the provided feedback types. They designed a labelling system to classify the feedback messages by using a established feedback taxonomy~\cite{narciss.2006,narciss.2008}. The following elaborate feedback types are relevant for the present study:
\begin{itemize}[leftmargin=*]
    \item \textbf{Knowledge about Task Constraints} Feedback that clarifies the problem's rules and requirements, such as reminding a student to use a specific language construct (Hints on task requirements) or suggesting a general approach to the problem (Hints on task-processing rules) ~\cite{keuning.2019}. 
    \item \textbf{Knowledge about Mistakes} Feedback that explicitly identifies an error. This can include pointing out Test failures, compiler errors, logic or runtime Solution errors, poor style issues, or suboptimal performance issues~\cite{keuning.2019}.
    \item \textbf{Knowledge about Concepts} Feedback that addresses the student's understanding of the subject matter, either through direct explanations of programming concepts or by providing illustrative examples~\cite{keuning.2019}.
    \item \textbf{Knowledge about How to Proceed} Constructive feedback that guides the student forward. This can be a bug-related hint on error correction, a suggestion for the next task-processing steps, or a hint for improvements to an already working solution ~\cite{keuning.2019}.
    \item \textbf{Knowledge about Meta-cognition} Feedback that prompts the student to reflect on their learning and problem-solving strategies, for instance, by asking about their confidence in a chosen approach or their plan for solving the task ~\cite{keuning.2019}.
\end{itemize}
\noindent
In addition, the \textbf{Strategic Processing Steps} type is suggested \cite{kiesler.2022}. This type provides knowledge on how to approach tasks strategically and systematically, enabling students to solve similar tasks or problems.
\citet{kiesler.2023c,kiesler.2022} had analyzed students' use of predefined tutoring feedback types~\cite{keuning.2019, narciss.2008} in an online programming tool. She concluded that learners need specific, task-related information to meet their cognitive and motivational needs. For educators, providing such feedback at scale was unfeasible until the emergence of GenAI. Earlier automated systems without GenAI relied mostly on static approaches (e.g., program repair \cite{kurniawan.2023}), output-matching and unit-tests for generating feedback, which often did not provide enough detail and could leave students frustrated~\cite{messer.2024}.

Advancements in GenAI have opened unprecedented potential for providing immediate feedback in programming education at scale -- even for diverse learners~\cite{becker.2023a, prather.2023, prather.2024c}. Recent studies demonstrate that models like GPT-4 can generate tiered~\cite{nguyen.2024} and various types~\cite{xiao.2024a, lohr.2025a} of feedback. Even though the quality of feedback messages generated by GPT-4 is improving, it is far from perfect~\cite{azaiz.2024, jacobs.2024, gabbay.2024}.
As a result, there is growing research on how to enhance the quality of the feedback messages with techniques like Retrieval Augmented Generation~\cite{jacobs.2024c, liu.2024a}, multiple prompts~\cite{birillo.2024, kazemitabaar.2024}, chain-of-thought prompting~\cite{zhu.2025}, fine-tuning ~\cite{solano.2025, liu.2025}, or validation of feedback with a smaller model~\cite{phung2023generativeaifor}. 
Recent research on feedback in computing education often focuses on the quality of feedback messages and how we can develop custom tools to support students. 
Yet, \citet{stone_exploring_2024} highlights a crucial gap in our understanding of how introductory programming students actually \textit{engage} with GenAI.

\subsection{Related Work on Feedback Engagement} \label{sec:Feedback-Engagement}
Regardless of the quality of a feedback message, \citet{mandouit.2023} emphasize the necessity for students to hear, understand, and apply the feedback they receive. \textit{Feedback engagement} can be described as students' perception, interpretation, and use of feedback~\cite{vanderkleij.2021}. \citet{price.2011} further characterizes the engagement process with (assessment) feedback on a temporal dimension. It starts with (1) students collecting feedback. Next, (2) students must devote their attention before they can (3) cogitatively engage with it. Based on their cognitive engagement and individual and situational factors, they might (4) take immediate or latent action.

In this perspective, learners have a more active role than that of a recipient who processes information (cognitive, motivational, meta-cognitive~\cite{narciss.2008} or cognitive, affective, and behavioral~\cite{lipnevich.2022}). \citet{winstone.2017} highlight this active role with the term \emph{proactive recipience}, accentuating the learner's responsibility for making feedback processes effective.
\citet{narciss.2008} describes an interaction between internal and external feedback loops: Learners evaluate external feedback against their own internal feedback, determining whether and how to act on the feedback. 
It is thus important to foster learners' feedback literacy~\cite{winstone.2019, cucuiat.2024}, that is the understanding, capacities, and dispositions required to make sense of the presented information and to use it successfully~\cite{carless.2018}.

\section{Methodology}

\begin{figure*}[htb]
    \centering
    \includegraphics[width=0.99\linewidth]{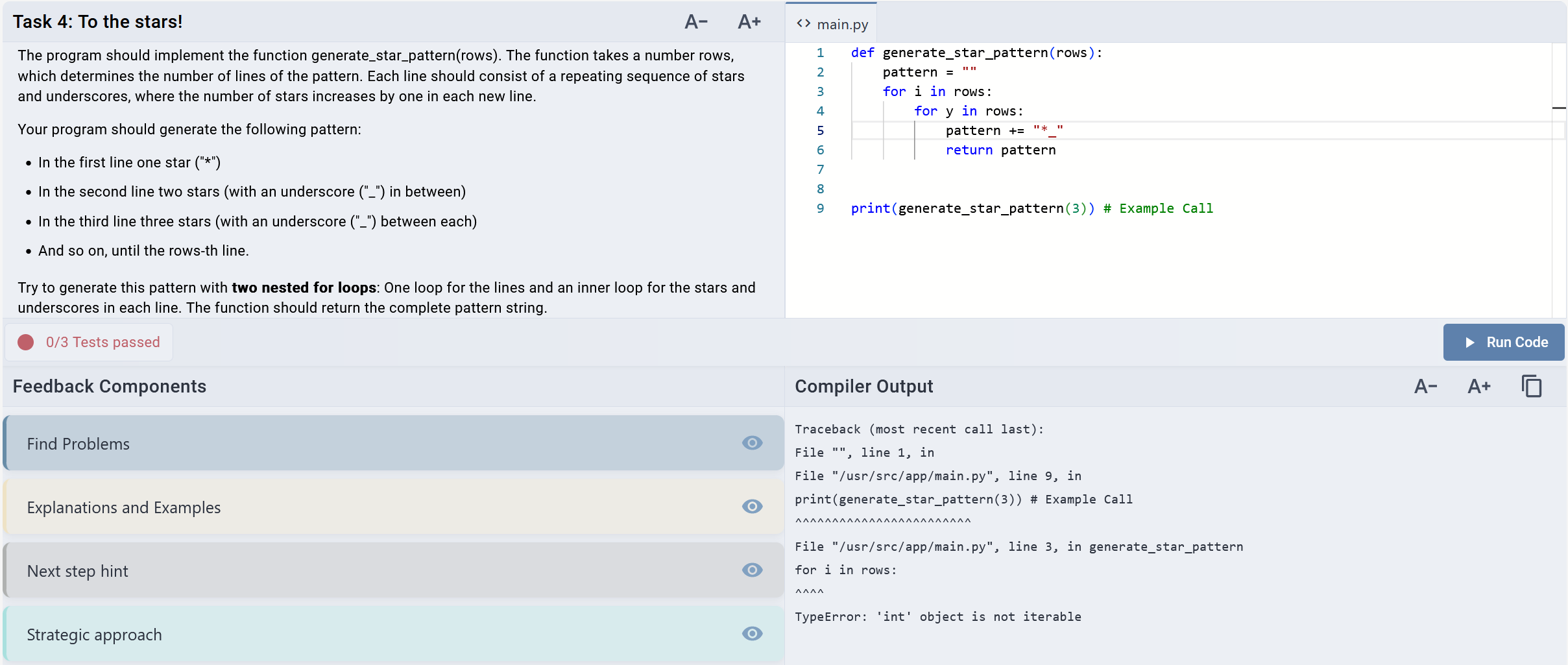}
    \caption{Code editor in the Tutor Kai (after a GenAI feedback was requested)}
    \label{fig:Code Editor}
\end{figure*}

To investigate students' thinking and engagement with different types of immediate tutoring feedback, we developed and provided four feedback components as part of the Tutor Kai~\cite{jacobs.2024, jacobs.2024c, jacobs.2025a, jacobs.2025, jacobs.2025c}.
Students' feedback requests, (visual) attention, cognitive engagement, and resulting actions~\cite{price.2011} are gathered and analyzed. We adopted a mixed-methods approach comprising eye-tracking and a think-aloud protocol, including a questionnaire. 

Eye-tracking provides the possibility to observe cognitive processing without adding to the cognitive load~\cite{busjahn.2014}. In this study, we use it to observe students' immediate (visual) attention when confronted with feedback. This allows us to see what students' eyes fixate~\cite{sharafi.2020} (e.g., the textual feedback). We interpret fixations on textual feedback as an indication that the feedback has been read.

The think-aloud protocol involves instructing participants to continuously express what goes through their minds and utter it without interpretation or explanation. These verbalizations during problem solving provide rich data for analyzing the involved cognitive processes~\cite{solomon.1994}. In this study, the method is applied to explore computing students' thinking while they solve programming tasks in an online learning environment. The eye-tracking data is collected simultaneously. 
We further used a questionnaire to collect demographic data (e.g., age, prior programming experience) and students' perspectives (e.g, perceived task difficulty).
Utilizing these methods and triangulating the resulting data is essential to gain insights into multiple processes of feedback engagement~\cite{price.2011}.

In the following, we present the feedback components as part of the Tutor Kai, as well as our data collection and analysis methods. The research data (e.g., programming tasks, think-aloud instructions, questionnaire, prompts used for GenAI feedback, examples of feedback components) is available online~\cite{supplementarydata.2025}.

\subsection{Feedback Components in the Tutor Kai} \label{sec:Feedback-Generation}

As a basis for this study, we provide students with a practice platform for Python programming exercises. It is the same platform they use for their weekly assignments as part of the Introduction to Programming class. Choosing this tool thus had several advantages, e.g., students were already familiar with it, and could sign on through their university's authentication. Moreover, we were able to provide distinct feedback components in a controlled environment through this tool, which is crucial for this study. 

The learning environment's UI comprises a task description (\autoref{fig:Code Editor}, top left) and a code editor (\autoref{fig:Code Editor}, top right). Students can execute their code as often as they wish, receiving compiler feedback (\autoref{fig:Code Editor}, bottom right). In addition, unit tests associated with the task are run automatically with each execution. Students can (re-)generate GenAI feedback by using a separate button, which is only visible after code changes have been executed (at least once).
GenAI feedback is provided via four components, which can be opened (i.e., collected) individually in any order (\autoref{fig:Code Editor}, bottom left).
The decision is based on findings in prior work~\cite{kiesler.2023c, jacobs.2024, jacobs.2025a}, suggesting a self-determined selection of feedback and its extent to foster the students' self-efficacy and competency. 

While the generation of isolated feedback types is possible with GenAI~\cite{lohr.2025a}, their strict separation can be ineffective in real-world applications \cite{kiesler.2023c}. We therefore developed the following four feedback components that represent various feedback types. 

The \emph{Find Problems (KM)} component describes students' mistakes. If the mistake relates to a compiler error, a brief explanation of the compiler message is provided. In addition, relevant information from the task description is supposed to help clarify the mistake. 

\begin{feedbackcard}{Find Problems}
    You tried to iterate directly over 'rows', but 'rows' is a number and not an iterable object. This leads to the error: 'TypeError: 'int' object is not iterable'. Additionally, the pattern is not built as required because the loops are not nested correctly and the 'return' is executed too early.
\end{feedbackcard}

The \emph{Explanations and Examples (KC)} component explains concepts relevant to the student's mistakes identified in the \emph{KM} component. It provides generic code examples for illustration and cites relevant lecture materials (video recordings from the lecture) when available.
This provides additional scaffolding opportunities, as the student is shown an exemplary implementation of a concept, which they have to adapt to their own problem.
The technique for linking to the precise moment in video lecture recordings where a concept is explained is implemented using Retrieval-Augmented Generation (RAG) and hypothetical student questions~\cite{jacobs.2024c}. Linked videos will be opened within a modal in the Tutor Kai.

\begin{feedbackcard}{Explanations and Examples}
    In Python, you can use a for loop to iterate over so-called 'iterable objects', such as lists or number sequences generated with 'range()'. If you want to iterate over a number, you must use 'range()' to generate a sequence of numbers (\textcolor{blue}{1}). Example:
    \begin{center}
    \begin{minipage}{0.9\linewidth}
    \begin{lstlisting}
    for i in range(5):
        print(i)
    \end{lstlisting}
    \end{minipage}
    \end{center}
    This outputs the numbers 0 through 4. So the loop runs five times. This is useful when you want to repeat something a specific number of times.
    
    \vspace{0.5em} 
    Sources (Links to Video):
    \begin{enumerate}[leftmargin=*, labelindent=0pt, itemindent=0pt]
        \item \textcolor{blue}{Python Control Structures}
    \end{enumerate}
\end{feedbackcard}

The \emph{Next step hint (KH)} component provides the student with an actionable next step to guide them toward correcting their mistakes (identified in KM). It can include a very short code snippet, but it does not provide the complete solution.

\begin{feedbackcard}{Next Step Hint}
    Use 'range()' in your for loop to count from 0 to rows-1. For example:
    \begin{center}
    \begin{minipage}{0.9\linewidth}
    \begin{lstlisting}
    for i in range(rows):
        # Your code for each line goes here
    \end{lstlisting}
    \end{minipage}
    \end{center}
    This allows you to generate the desired number of lines.
\end{feedbackcard}

The \emph{Strategic Approach (SPS)} component provides a generally formulated, step-by-step description of procedures for solving tasks of a similar nature. We do not provide the correct solution, but strategic processing steps~\cite{kiesler.2022} to prevent trial-and-error behavior. 

\begin{feedbackcard}{Strategic Approach} 
    To solve tasks like this where a specific pattern needs to be created with nested loops, you can apply the following strategy:
    \begin{enumerate}[leftmargin=*, labelindent=0pt, itemindent=0pt]
        \item Consider\,how\,many\,lines\,the\,pattern\,should\,have\,and\,how\,you\,generate\,each\,line.
        \item Use an outer loop that runs once for each line (e.g., from 1 to rows).
        \item Use an inner loop to generate the stars and underscores in the current line. The number of stars corresponds to the current line number.
        \item Make\,sure\,that\,underscores\,appear\,between\,stars,\,but\,not\,at\,the\,end\,of\,the\,line.
        \item Add a line break (\textbackslash n) after each line (except the last one).
        \item Test your program with different inputs to ensure the pattern is generated correctly.
    \end{enumerate}
    This method is effective because it breaks down the problem into small, manageable steps, and nested loops are a typical solution for such pattern tasks.
\end{feedbackcard}

We used OpenAI's \texttt{gpt-4.1-2025-04-14} model with a temperature setting of 0 and structured output for feedback generation in all components. As context, we provided the task description, students' code, corrected version of students' code, relevant lecture transcripts, compiler output, unit tests, and their results. A full system description, including all prompts, example inputs, and outputs, is available in our supplementary data repository~\cite{supplementarydata.2025}.

\subsection{Data Collection} \label{sec:Data-Collection}

\subsubsection{Task Design} \label{sec:Task-Design}
Six tasks were selected in alignment with the course progress and students' expected proficiency at the time of the study (see \autoref{tab:tasks}). These tasks had been used in a prior study~\cite{jacobs.2025a} in this course before, were only available to participants, and were tested by two tutors in the same environment.
The complete tasks, including model solutions and unit tests, are available in our supplementary data repository~\cite{supplementarydata.2025}.

\begin{table}[thb]
  \centering
  \footnotesize
  \caption{Tasks}
  \label{tab:tasks}
  \begin{tabular}{@{}
    p{0.03\textwidth}
    p{0.42\textwidth}
  @{} }
    \toprule
    \textbf{Task}& \textbf{Concepts} \\
    \midrule
    T1 & Function, Parameter, Adding and multiplying variables, return\\
    T2 & Function, Parameter, IF-Else control structure, comparison operators, return\\
    T3 & Function, Parameter, If-Else control structure, comparison operators, loop, list, append to list, return\\
    T4 & Function, Parameter, If-Else control structure, comparison operators, nested loops, sequences of numbers with range, string concatenation, return  \\
    T5 & Function, Parameter, If-Else control structure, comparison operators, recursion, return \\
    T6 & Function, Parameter, Adding and multiplying variables, Higher-Order Functions, return \\
    \bottomrule
  \end{tabular}
\end{table}

\subsubsection{Questionnaire Design}
A questionnaire was designed to gather students' programming experience, their perception of the task difficulty, and their ratings of the helpfulness of the feedback components.
Students rated their programming experience on a Likert scale from 1 (very inexperienced) to 10 (very experienced). We also asked students to provide their programming experience in years, and how many programming-related courses they have already taken at the university.
Furthermore, the perceived difficulty of each task was rated on a 5-point Likert scale from 1 (the task was very easy for me) to 5 (the task was very difficult for me).
We additionally invited students to describe any positive or negative effects they experienced from the think-aloud protocol and from wearing eye-tracking glasses via two open input fields. 
The full questionnaire is available in our supplementary data repository~\cite{supplementarydata.2025}.

\subsubsection{Participants}
The study was conducted in an introductory programming course at a German university. The target group of students is in their first or second semester. The data collection took place during the third and fourth weeks of the summer semester 2025. We invited all (appr. 200) students by mail and in person during the lecture in the first week of the semester. We offered them compensation of 40€, but participation was completely voluntary and had no impact on students' grades or success in the course.

\subsubsection{Setting and Procedure}
Students used standard input (mouse, keyboard) and output devices (27-inch monitor, headphones) while solving the tasks.
For eye tracking, we used the Tobii Glasses 3, together with the Tobii Pro Lab analysis software.
A two-hour time slot was reserved for each participant. The procedure within each time slot was structured as follows:

\begin{enumerate}[leftmargin=*]
    \item Introduction (20-40 minutes):
    \begin{itemize}
        \item Informed consent process
        \item Think-aloud protocol instruction and practice, including one demonstration task explained by the instructor and one practice task completed by the participant
        \item Eye-tracking equipment familiarization and calibration
    \end{itemize}
    \item Task Phase (60 minutes):
    \begin{itemize}
        \item Completion of programming tasks
        \item Think-aloud verbalization
        \item Eye-tracking and screen recording
        \item Video and audio recording from front webcam
    \end{itemize}
    \item Questionnaire (10-20 minutes)
\end{enumerate}

The informed consent process includes a full disclosure of all technical equipment, the data being recorded, the data analysis, and the rights of the participants. All participants gave their written consent. 
For technical inquiries, the same instructor was always present during data collection.

\subsection{Data Analysis}

\subsubsection{Data Preparation}
First, the recordings from the Tobii Glasses 3 were exported using the Tobii Pro Lab (25.7) analysis software. This process generates a video that overlays the gaze visualization onto the video recording from the glasses.
We used the Tobii I-VT (fixation) gaze filter algorithm~\cite{olsen.2012} with default settings (minimum fixation duration of 60 milliseconds).
This video allows us to see where the student is looking at any given moment. We then synchronized this video with the screen recording and the webcam recording to create a single \emph{synchronized video} consisting of three synchronized video streams. We used the audio recording from the Tobii Glasses 3, as students' verbalized thoughts were clearest on this track. In the next step, transcripts for the synchronized videos were generated locally on our machines using the open-weight whisper-large-v3-turbo speech recognition model~\cite{radford.2022}. We imported the synchronized videos and their corresponding transcripts into MAXQDA for qualitative data analysis and coding~\cite{kuckartz.2019}.

We manually corrected the generated transcripts. Additionally, we augmented the transcripts with data from the database logs.
Whenever a student executed their code using ``Run Code'', the corresponding database entry containing the student's source code and associated compiler output was inserted into the transcript and coded as such. If the student's solution was correct, we coded it as a successful task completion. 
When a student opened a GenAI feedback component, we inserted the corresponding feedback text into the transcript. We also coded the four feedback components as such (see \autoref{sec:Feedback-Generation}) and whether they were read or ignored.

Visual attention was analyzed using gaze visualizations available within the synchronized videos. This allowed us to observe and follow students' fixations (visualized as a red dot) at every moment during the recording. We interpret zero fixations on feedback component texts as indicating that they were not read (ignored). As we have no insights into each student's individual reading behavior, we interpret multiple fixations on a feedback component as students having (partially) read this feedback component. 

\subsubsection{RQ1: Think-Aloud Analysis}
We investigated what students thought (aloud) during engagement with a feedback component via MAXQDA. We imported the synchronized video and its transcript, augmented with data from the database logs. Using both the synchronized video and its transcript was important for data triangulation. The gaze visualization, as part of the synchronized video, lets us observe the feedback component that students relate to when verbalizing thoughts. We mapped this relation based on (1) the feedback the student most recently looked at (gaze visualization), (2) the currently collected component (screen recording), and (3) the semantics of verbalized thoughts and the feedback component.

Students' verbalized thoughts (ranging from a single word to multiple sentences) related to a feedback component constitute a coding unit. If in doubt, we utilized the synchronized video and its transcript for the same programming task as a context unit.
To ensure completeness, we had added the texts of the feedback components during the data preparation phase. 
We also used MAXQDA's~\cite{kuckartz.2019} quantitative capabilities for analyzing frequencies and co-occurrences between feedback components, verbalized thoughts, and actions. 

Two of the authors coded the data inductively. They met regularly to discuss emerging codes and develop a preliminary category system. This occurred at increasing intervals, starting with two meetings for every 10\% (i.e., material of two students) and then several times for every 20\% of the material. Codes were thematically grouped to identify overarching patterns for analysis. For example, students thought aloud questions regarding information in the feedback component were grouped based on the types of difficulties~\cite{qian.2018} these questions addressed. The third author was consulted on unclear cases to reach a consensus~\cite{spangler2012consensual}.
This iterative process led to the construction of the final coding scheme, which defined each category with accompanying definitions and anchor examples. 
When saturation was achieved (i.e., no new codes emerged) within the sample, the final coding scheme was applied to the entire material. 

\subsubsection{RQ2: Immediate Actions and Outcome Analysis}

To answer RQ2, we developed a deductive coding scheme for students' immediate actions based on the observable interactions and outcomes. The coding schema (code changes, feedback collection, or no action) is based on the possible feedback-related actions within the learning environment (see \autoref{sec:Feedback-Generation}).

Actions were observable within the screen recording part of the synchronized videos. For code changes, we used the final student's code state related to a feedback component as the coding unit. All prior logs related to the same task (added to transcripts during data preparation) served as a context unit. Other actions, like collecting another feedback component, were objectively based on logs.

We mapped students' actions to feedback components based on (1) the feedback component they most recently read, (2) the semantics of code changes made and information in the feedback text, and (3) their co-occurrent (verbalized) thoughts. Instances where action was lacking were coded accordingly.

To investigate the relationships between the feedback components, students' (verbalized) thoughts, and their (immediate) actions, we used  MAXQDA's so-called Complex Coding Query function. It analyzes the co-occurrence of multiple codes (feedback component, thoughts, and actions) within the same data segment.

An inter-coder reliability test for coded attention, thoughts, and actions was conducted. A random sample of three participants (15\% of the data) was coded independently by authors one and two. Using MAXQDA, we analyzed segment-level agreement and calculated a Kappa coefficient \cite{brennan.1981}. The analysis yielded a substantial \cite{landis.1977, kuckartz.2019} inter-coder reliability of $ \kappa_n = 0.70$.

\begin{table*}[thb]
    \setlength{\tabcolsep}{2.2pt}
    \centering
    \footnotesize
    \caption{Participants}
    \label{tab:participants}
    \begin{tabular}{@{}l*{21}{c}@{}}
    \toprule
    \textbf{Student} & \textbf{S1} & \textbf{S2} & \textbf{S3} & \textbf{S4} & \textbf{S5} & \textbf{S6} & \textbf{S7} & \textbf{S8} & \textbf{S9} & \textbf{S10} & \textbf{S11} & \textbf{S12} & \textbf{S13} & \textbf{S14} & \textbf{S15} & \textbf{S16} & \textbf{S17} & \textbf{S18} & \textbf{S19} & \textbf{S20} & \textbf{M (SD)*} \\
    \midrule
    \multicolumn{21}{@{}l}{\textbf{Background Information}} \\
    \addlinespace[2pt]
    Self-reported experience (1-10) & 2 & 2 & 4 & 1 & 3 & 4 & 6 & 3 & 1 & 8 & 6 & 5 & 5 & 2 & 5 & 4 & 4 & 2 & 1 & 7 & 3.8 (2.1)\\
    Years of programming & 0.5 & 0 & 3 & 1 & 2 & 1 & 6 & 3 & 0.3 & 10 & 3 & 2 & 1 & 0 & 1 & 1 & 3 & 0 & 1 & 9 & 2.4 (2.8)\\
    Courses attended & 1 & 1 & 2 & 1 & 1 & 2 & 2 & 4 & 0 & 2 & 1 & 2 & 1 & 2 & 1 & 0 & 2 & 1 & 2 & 3 & 1.6 (0.9)\\
    \midrule
    \multicolumn{21}{@{}l}{\textbf{Feedback Usage (Count)}} \\
    \addlinespace[2pt]
    GenAI feedback requests & 15 & 9 & 6 & 4 & 7 & 12 & 16 & 11 & 12 & 2 & 10 & 7 & 4 & 13 & 13 & 8 & 6 & 7 & 4 & 6 & 8.6 (4.0)\\
    \hspace{0.5em}Find Problems (KM) & 15 & 9 & 6 & 4 & 7 & 9 & 16 & 9 & 12 & 1 & 10 & 7 & 4 & 13 & 13 & 8 & 6 & 7 & 4 & 6 & 8.3 (4.0)\\
    \hspace{0.5em}Explanations and Examples (KC) & 6 & 8 & 4 & 2 & 5 & 11 & 12 & 6 & 7 & 1 & 6 & 7 & 2 & 8 & 11 & 3 & 5 & 4 & 2 & 4 & 5.7 (3.2)\\
    \hspace{0.5em}Next Step Hint (KH) & 6 & 6 & 4 & 1 & 2 & 8 & 11 & 8 & 1 & 0 & 2 & 5 & 1 & 3 & 6 & 2 & 3 & 3 & 1 & 1 & 3.7 (3.0)\\
    \hspace{0.5em}Strategic approach (SPS) & 3 & 3 & 1 & 1 & 1 & 4 & 2 & 2 & 0 & 1 & 1 & 3 & 0 & 3 & 2 & 0 & 0 & 1 & 1 & 1 & 1.5 (1.2)\\
    \midrule
    
    \multicolumn{21}{@{}l}{\textbf{Perceived Task Difficulty (1-5) and Task Performance (Y: solved, N: not solved)}} \\
    \addlinespace[2pt]
    T1 & 2 / Y & 1 / Y & 1 / Y & 5 / Y & 1 / Y & 2 / Y & 1 / Y & 2 / Y & 2 / Y & 1 / Y & 1 / Y & 1 / Y & 1 / Y & 2 / Y & 1 / Y & 1 / Y & 1 / Y & 4 / Y & 2 / Y & 1 / Y & 1.7 (1.1)\\
    T2 & 2 / Y & 2 / Y & 2 / Y & 5 / Y & 1 / Y & 2 / Y & 2 / Y & 3 / Y & 3 / Y & 1 / Y & 1 / Y & 2 / Y & 1 / Y & 2 / Y & 1 / Y & 1 / Y & 2 / Y & 4 / Y & 2 / Y & 1 / Y & 2.0 (1.1)\\
    T3 & 4 / N & 5 / N & 2 / Y & 3 / Y & 4 / N & 1 / N & 2 / Y & 4 / Y & 4 / Y & 1 / Y & 2 / Y & 2 / Y & 1 / Y & -- / Y & 3 / Y & 4 / Y & 3 / Y & 5 / N & 2 / Y & 2 / Y & 2.8 (1.3)\\
    T4 & 4 / N & 5 / N & 4 / N & 1 / N & -- / N & -- / N & 5 / N & 5 / N & 5 / N & 1 / Y & 4 / N & 5 / N & 3 / Y & 5 / N & 5 / N & 5 / N & -- / N & -- / N & 4 / N & 4 / N & 4.1 (1.3)\\
    T5 & -- / N & -- / N & -- / N & -- / N & -- / N & -- / N & -- / N & -- / N & -- / N & 4 / Y & -- / N & -- / N & 4 / Y & 5 / N & -- / N & -- / N & -- / N & -- / N & -- / N & -- / N & 4.3 (0.6)\\
    T6 & -- / N & -- / N & -- / N & -- / N & -- / N & -- / N & -- / N & -- / N & -- / N & 1 / Y & -- / N & -- / N & 1/ Y & -- / N & -- / N & -- / N & -- / N & -- / N & -- / N & -- / N & 1.0 (0.0)\\
    \bottomrule
    \addlinespace[1pt]
    \multicolumn{20}{@{}l@{}}{\makecell[l]{\footnotesize *M = Mean, SD = Standard Deviation}} \\
    \end{tabular}
\end{table*}

\begin{table}[h]
\setlength{\tabcolsep}{2pt}
  \centering
  \footnotesize
  \caption{Feedback components and (visual) attention}
  \label{tab:feedback_usage}
    \begin{tabular}{@{}
    p{0.2\textwidth}
    p{0.06\textwidth}
    p{0.06\textwidth}
    p{0.06\textwidth}
    p{0.06\textwidth}
    @{} }
    \toprule
    \textbf{Feedback Components} & \textbf{Generated} & \textbf{Collected} & \textbf{Read} & \textbf{Ignored} \\
    \midrule

    \addlinespace[1pt]
    Find Problems (KM)
       & 172
      & 166
      & 166
      & 0 \\
    \addlinespace[1pt]
    Explanations and Examples (KC)
    & 172
      & 114
      & 111
      & 3 \\
    \addlinespace[1pt]
    Next Step Hint (KH)
    & 172
      & 74
      & 73
      & 1 \\
    \addlinespace[1pt]
    Strategic approach (SPS)
    & 172
      & 30
      & 30
      & 0 \\
    \bottomrule
     \addlinespace[1pt]
  \end{tabular}
\end{table}

\section{Results}

In this section, we present the results, starting with a short summary of the overall student sample. Moreover, we provide the number of student feedback requests and visual attention.  

\textit{Student Sample:} A total of 23 students participated in the study. Due to technical issues and individual health conditions, precise eye-tracking with the Tobii Glasses 3 was not possible for three participants. So we only analyzed the data from the remaining 20 students. 15 students identified as male and 5 as female. The mean age was 22.45 years (SD = 2.97). 
Students self-reported diverse programming experience on a 5-point scale (from 1 = ``very inexperienced'' to 10 = ``very experienced'') with an average of 3.8 (SD = 2.1). This diversity is notable in their years of programming experience and the number of programming-related courses taken (see \autoref{tab:participants}).
The students perceived the tasks as becoming increasingly difficult. On a 5-point scale (from 1 = ``very easy for me'' to 5 = ``very difficult for me''). The average rating increased from 1.7 for Task 1 to 4.3 for Task 5 (see \autoref{tab:participants}). On average, students solved 3.05 of 6 tasks (SD = 1.07).

\textit{Students' feedback requests and visual attention:}
GenAI-feedback was requested 172 times (see \autoref{sec:Feedback-Generation}). Since all four feedback components were generated each time, a total of 688 feedback components were generated. Students collected 384 (55.8\%) of these 688 components (see \autoref{tab:feedback_usage}). By analyzing the gaze videos, we observed that 380 of the 384 collected feedback components were (at least partially) read.

\subsection{RQ 1: Students' Verbalized Thoughts during Feedback Engagement} \label{sec:results:thoughts}

\begin{table*}[thb]
\setlength{\tabcolsep}{2pt}
  \centering
  \footnotesize
  \caption{Coding scheme for students' thoughts when engaging with tutoring feedback components}
  \label{tab:thoughts}
  \begin{tabular}{@{}
    p{0.12\textwidth}
    p{0.28\textwidth}
    p{0.35\textwidth}
    c c c c c
  @{} }
    \toprule
    \textbf{Category} & \textbf{Definition} & \textbf{Anchor Examples} & \multicolumn{5}{c}{\textbf{Occurrences}} \\
    \cmidrule(lr){4-8}
    & & & \textbf{Total} & \textbf{KM} & \textbf{KC} & \textbf{KH} & \textbf{SPS} \\
    \midrule
    \multicolumn{3}{@{}l@{}}{\makecell[l]{\textbf{Students express understanding:}}} \\
    
    \addlinespace[1pt]
    Agrees to Information
      & Expresses nonspecific agreement with the statement or information given.
      & \say{Ah, okay}; \say{Hmm, great} 
      & 89
      & \makecell{49\\(29.5\%)}
      & \makecell{19\\(17.1\%)}
      & \makecell{17\\(23.3\%)}
      & \makecell{4\\(13.3\%)}
      \\

    \addlinespace[1pt]
    Thoughts demonstrate understanding
      & Articulated thoughts imply that the feedback has been understood (e.g., by verbalizing a necessary code change).
      & \say{Maybe I have to combine for and if somehow}; \say{So I don't need the index anymore.}
      & 79
      & \makecell{38\\(22.9\%)}
      & \makecell{25\\(22.5\%)}
      & \makecell{12\\(16.4\%)}
      & \makecell{4\\(13.3\%)}
      \\

    \addlinespace[1pt]
    Restates feedback in own words
      & Restates (parts) of the given information in their own words.
      & \say{That means, if we are dealing with a number, or rather an integer, we have to use 'range' in the for loop.}; \say{okay, good, exactly, it's suggesting that I just use if-else here}
      & 17
      & \makecell{11\\(6.6\%)}
      & \makecell{3\\(2.7\%)}
      & \makecell{3\\(4.1\%)}
      & \makecell{0\\(0.0\%)}
      \\

    \midrule
    \multicolumn{3}{@{}l@{}}{\makecell[l]{\textbf{Additional information needed:}}} \\
    
    \addlinespace[1pt]
    I don't understand
      & Does not understand a piece of information, without specifying why.
      & \say{I don't understand.}; \say{But I didn't manage to understand exactly how this case is incorrect.}
      & 8
      & \makecell{5\\(3.0\%)}
      & \makecell{1\\(0.9\%)}
      & \makecell{1\\(1.4\%)}
      & \makecell{1\\(3.3\%)}
      \\
    
    \addlinespace[1pt]
    Syntax and Implementation
      & Is uncertain about a syntactic detail. Knows what to do, but not how to implement it.
      & \say{I really don't know how to incorporate that right now.}; \say{How do you calculate a product now?}
      & 38
      & \makecell{22\\(13.3\%)}
      & \makecell{6\\(5.4\%)}
      & \makecell{5\\(6.9\%)}
      & \makecell{5\\(16.7\%)}
      \\

    \addlinespace[1pt]
    Concepts and Explanations
      & Requires additional explanation or is uncertain about a programming concept.
      & \say{What was 'append' again?}; \say{This makes note a string and not a numerical value. Whoa, what exactly is meant by that now?}
      & 16
      & \makecell{9\\(5.4\%)}
      & \makecell{2\\(1.8\%)}
      & \makecell{5\\(6.9\%)}
      & \makecell{0\\(0.0\%)}
      \\

    \addlinespace[1pt]
    Strategy and next steps
      & Does not know what to do next. For example, having problems understanding the task or creating a strategy to solve it.
      & \say{Because I don't know how to proceed.}; \say{I have no idea what I'm supposed to do here.}
      & 26
      & \makecell{10\\(6.0\%)}
      & \makecell{8\\(7.2\%)}
      & \makecell{0\\(0.0\%)}
      & \makecell{8\\(26.7\%)}
      \\

    \addlinespace[1pt]
    \midrule
    
    \multicolumn{3}{@{}l@{}}{\makecell[l]{\textbf{Students express disagreement:}}} \\
    
    \addlinespace[1pt]
    Disagree
      & Contradicts given information or questions its correctness.
      & \say{But grades are the score, aren't they?}; \say{Although, that wouldn't really make much sense.}
      & 12
      & \makecell{5\\(3.0\%)}
      & \makecell{1\\(0.9\%)}
      & \makecell{5\\(6.9\%)}
      & \makecell{1\\(3.3\%)}
      \\

    \addlinespace[1pt]
    Irrelevant to the current problem
      & States that the information does not match the problem currently worked on.
      & \say{But that still doesn't really solve the problem with the comparison operators.}; \say{It's not very helpful that it focuses so much on the return statement, as that isn't the main problem for me.}
      & 25
      & \makecell{5\\(3.0\%)}
      & \makecell{9\\(8.1\%)}
      & \makecell{8\\(11.0\%)}
      & \makecell{3\\(10.0\%)}
      \\
      
    \addlinespace[1pt]
    Information already known
      & Already knows the information (e.g., previous GenAI feedback or prior knowledge).
      & \say{Yes, unfortunately, I already know that.}; \say{This is mostly the same feedback I received before.} 
      & 39
      & \makecell{9\\(5.4\%)}
      & \makecell{20\\(18.0\%)}
      & \makecell{9\\(12.3\%)}
      & \makecell{1\\(3.3\%)}
      \\

    \addlinespace[1pt]
    \midrule

    \multicolumn{3}{@{}l@{}}{\makecell[l]{\textbf{Students' explicit judgments:}}} \\
    \addlinespace[2pt]

    Helpful
      & Judges the information helpful.
      & \say{Yes, that helps me.}; \say{This hint here was also helpful, that an integer isn't a list.}
      & 26
      & \makecell{4\\(2.4\%)}
      & \makecell{10\\(9.0\%)}
      & \makecell{7\\(9.6\%)}
      & \makecell{5\\(16.7\%)}
      \\

    \addlinespace[1pt]
    Not Helpful
      & Judges the information not helpful.
      & \say{Yeah, this is no help to me either.}; \say{Unfortunately, that doesn't help me.}
      & 32
      & \makecell{9\\(5.4\%)}
      & \makecell{10\\(9.0\%)}
      & \makecell{7\\(9.6\%)}
      & \makecell{6\\(20.0\%)}
      \\
      
    \bottomrule
    \addlinespace[1pt]
    \multicolumn{8}{@{}l@{}}{\makecell[l]{\footnotesize Percentages show proportion relative to total read feedback components (KM: n=166, KC: n=111, KH: n=73, SPS: n=30).}} \\
  \end{tabular}
\end{table*}

To answer RQ1, we analyzed the thoughts during the think-aloud protocol for 380 feedback components that were actively collected and read by students (see \autoref{tab:feedback_usage}). We were able to categorize verbalized thoughts for 314 of them. Precisely, we identified 12 categories (see \autoref{tab:thoughts}), and assigned them to four main themes: \emph{Students express understanding} (45.5\% of verbalized thoughts), \emph{Additional information needed} (21.6\%), \emph{Students express disagreement} (18.7\%), and \emph{Students' explicit judgments} (14.3\%).

For the remaining 66 (out of 380 collected and read) feedback components, we were unable to define categories as students' thoughts were highly diverse (in 23 cases), or not uttered at all (in 43 cases). For example, only one student thought aloud \say{Well, I feel like I cheated a bit} after the \emph{next step hint} feedback component provided one line of code, which was key for solving T5.

\subsubsection{Students Express Understanding}

\textit{} \newline \textit{Agrees to Information:} Students expressed agreement in 89 cases. This often occurred immediately after or while reading individual pieces of information within a feedback component. However, a student sometimes initially agreed with a piece of information, and then expressed disagreement at a later stage. The opposite case was also observed, e.g., a student read the first half of the \emph{Explanations and Examples (KC)} feedback component, which explains comparison operators using an example. Then they express \emph{this is irrelevant to the current problem} (see \autoref{sec:results:thoughts}). Still reading the same feedback component, the student sees the explanation for the append method and says, \say{Ah! Append. Okay yes.}. For this reason, there are 20 co-occurrences of categories from \emph{additional information needed} and 2 co-occurrences from \emph{students express disagreement} within the same coding units.

\textit{Thoughts demonstrate understanding:} In 79 cases, students' articulated thoughts implying the feedback was understood. This was typically observed through the verbalization of required code changes or an explicit reflection of an existing error. In four cases, there were co-occurrences with \emph{Syntax and Implementation}, meaning additional information was needed. 
For example, a student read a \emph{Find Problems (KM)} feedback component for Task T3. It states that the comparison operator in their code, \texttt{=<}, does not exist, without specifying the correct syntax (this would be part of the \emph{KH} feedback component). The student stated that they understand the error description, but are unsure of how to correct their code. As the feedback had only described the error and provided no information on its correction, the students' thoughts still demonstrate that the feedback component itself was understood.

\textit{Restates feedback in own words:} Students restated information from the feedback component in their own words in 17 cases. 7 of these 17 cases originated from a single student (S3). Typically, the student would first read the complete feedback and then repeat information from that feedback component in their own words, thereby revealing understanding. There were 5 co-occurrences with \emph{Syntax and Implementation} in the same coding unit. Four of them involved \emph{Find Problems (KM)} feedback components. In these instances, students asked further questions regarding implementation details (see \autoref{sec:Feedback-Generation}), implying understanding.

\begin{table*}[t]
\setlength{\tabcolsep}{3.7pt}
  \centering
  \footnotesize
  \caption{Coding Scheme representing students' actions after reading tutoring feedback components}
  \label{tab:actions}
    \begin{tabular}{@{}
    p{0.12\textwidth}
    p{0.2\textwidth}
    p{0.38\textwidth}
    c c c c c
  @{} }
    \toprule
    \textbf{Category} & \textbf{Definition} & \textbf{Anchor Examples} & \multicolumn{5}{c}{\textbf{Occurrences}} \\
    \cmidrule(lr){4-8}
    & & & \textbf{Total} & \textbf{KM} & \textbf{KC} & \textbf{KH} & \textbf{SPS} \\
    \midrule
    
    IMP: Improvement
      & Changes to code related to the given information, getting closer to a correct solution. 
      & The student reads that a loop is necessary. He improves his code and implements the required for-loop. In doing so, he makes a small syntax error. Nevertheless, he has come closer to a correct solution overall. 
      & 160
      & \makecell{69\\(41.6\%)}
      & \makecell{42\\(37.8\%)}
      & \makecell{38\\(52.1\%)}
      & \makecell{11\\(36.7\%)}
      \\
      
    SET: Setback
      & Changes to code related to the given information, moving further away from a correct solution.
      & The student initially iterates over fixed values using the range function. Feedback suggests iterating based on the variable \texttt{rows} (number). The student implements \texttt{for i in rows}.
      & 13
      & \makecell{6\\(3.6\%)}
      & \makecell{2\\(1.8\%)}
      & \makecell{3\\(4.1\%)}
      & \makecell{2\\(6.7\%)}
      \\

    BOT: Improvement and Setback
      & Both positive and negative changes to code related to the given information.
      & Based on the feedback, the student uses a for-loop (Improvement), but changes variables elsewhere, which leads to a new problem (Setback). 
      & 28
      & \makecell{10\\(6.0\%)}
      & \makecell{12\\(10.8\%)}
      & \makecell{4\\(5.5\%)}
      & \makecell{2\\(6.7\%)}
      \\

    NOA: No Action (in Code)
      & No changes to the program code, or changes that have no connection to the given information.
      & Student reads feedback but continues exploring other approaches or decides to collect an additional feedback component.
      & 179
      & \makecell{79\\(47.6\%)}
      & \makecell{56\\(50.0\%)}
      & \makecell{29\\(39.7\%)}
      & \makecell{15\\(50.0\%)}
      \\

    COF: Collect another Feedback Component
      & Decides to collect an additional feedback component
      & After reading KM, the student opens KC for more details.
      & 212
      & \makecell{115\\(69.3\%)}
      & \makecell{70\\(63.1\%)}
      & \makecell{23\\(31.5\%)}
      & \makecell{4\\(13.3\%)}
      \\

    \bottomrule
    \addlinespace[1pt]
    \multicolumn{8}{@{}l@{}}{\makecell[l]{\footnotesize Percentages show proportion relative to total read feedback components (KM: n=166, KC: n=111, KH: n=73, SPS: n=30).}} \\
  \end{tabular}
\end{table*}

\begin{table*}[htb]
    \caption{Verbalized thoughts in relation to the read feedback components and immediate actions} 
    \label{tab:thoughts-feedback-action-outcome-relation}
    \begin{center}
    \footnotesize
    \setlength{\tabcolsep}{1.65pt}
    \begin{tabular}{l|cccccc|cccccc|cccccc|cccccc}
    \toprule
     \textbf{Feedback Components} & \multicolumn{6}{c|}{\textbf{KM} {\smaller n=139}} & \multicolumn{6}{c|}{\textbf{KC} {\smaller n=85}} & \multicolumn{6}{c|}{\textbf{KH} {\smaller n=63}} & \multicolumn{6}{c}{\textbf{SPS} {\smaller n=27}} \\
    \arrayrulecolor{gray!60}\midrule\arrayrulecolor{black} 
    Thoughts \hfill \textbackslash{} \hfill Actions & Total & IMP & SET & BOT & NOA & COF & Total & IMP & SET & BOT & NOA & COF & Total & IMP & SET & BOT & NOA & COF & Total & IMP & SET & BOT & NOA & COF \\
    \midrule 

    \multicolumn{1}{l|}{\emph{Students express understanding:}} & \multicolumn{6}{c|}{} & \multicolumn{6}{c|}{} & \multicolumn{6}{c|}{} & \multicolumn{6}{c}{} \\
    Agrees to Information & 49 & 31 & 3 & 3 & 12 & 29     & 19 & 12 & 0 & 3 & 4 & 8   & 17 & 13 & 3 & 0 & 1 & 2    & 4 & 3 & 0 & 1 & 0 & 0 \\
    Thoughts demonstrate understanding & 38 & 32 & 0 & 2 & 4 & 16      & 25 & 23 & 0 & 1 & 1 & 7   & 12 & 12 & 0 & 0 & 0 & 1    & 4 & 4 & 0 & 0 & 0 & 0 \\
    Restates feedback in own words & 11 & 5 & 0 & 1 & 5 & 7        & 3  & 3  & 0 & 0 & 0 & 2   & 3  & 3  & 0 & 0 & 0 & 0    & 0 & 0 & 0 & 0 & 0 & 0 \\
    \midrule
    \multicolumn{1}{l|}{\emph{Additional information needed:}} & \multicolumn{6}{c|}{} & \multicolumn{6}{c|}{} & \multicolumn{6}{c|}{} & \multicolumn{6}{c}{} \\
    I don't understand & 5 & 0 & 0 & 0 & 5 & 5         & 1  & 0 & 0 & 0 & 1 & 1    & 1 & 0 & 0 & 1 & 0 & 1      & 1 & 0 & 0 & 0 & 1 & 0 \\
    Syntax and Implementation & 22 & 7 & 1 & 1 & 13 & 19      & 6 & 3 & 0 & 0 & 3 & 5     & 5 & 2 & 0 & 1 & 2 & 3      & 5 & 2 & 1 & 1 & 1 & 0 \\
    Concepts and Explanations & 9 & 4 & 0 & 0 & 5 & 7         & 2 & 0 & 1 & 1 & 0 & 2     & 5 & 4 & 0 & 0 & 1 & 1      & 0 & 0 & 0 & 0 & 0 & 0 \\
    Strategy and next steps & 10 & 2 & 2 & 0 & 6 & 8        & 8 & 0 & 0 & 0 & 8 & 7     & 0 & 0 & 0 & 0 & 0 & 0      & 8 & 1 & 0 & 0 & 7 & 3 \\
    \midrule
    \multicolumn{1}{l|}{\emph{Students express disagreement:}} & \multicolumn{6}{c|}{} & \multicolumn{6}{c|}{} & \multicolumn{6}{c|}{} & \multicolumn{6}{c}{} \\
    Information already known & 9 & 0 & 0 & 0 & 9 & 9         & 20 & 1 & 0 & 2 & 17 & 17  & 9 & 0 & 0 & 0 & 9 & 2      & 1 & 0 & 0 & 0 & 1 & 0 \\
    Irrelevant to the current problem         & 5 & 1 & 0 & 0 & 4 & 4         & 9 & 2 & 1 & 1 & 5 & 7     & 8 & 1 & 0 & 1 & 6 & 3      & 3 & 2 & 0 & 0 & 1 & 0 \\
    Disagree & 5 & 0 & 1 & 0 & 4 & 3         & 1 & 1 & 0 & 0 & 0 & 0     & 5 & 0 & 0 & 1 & 4 & 4      & 1 & 0 & 0 & 0 & 1 & 0 \\
    \midrule
    \multicolumn{1}{l|}{\emph{Students' explicit judgments:}} & \multicolumn{6}{c|}{} & \multicolumn{6}{c|}{} & \multicolumn{6}{c|}{} & \multicolumn{6}{c}{} \\
    Helpful     & 4 & 3 & 0 & 1 & 0 & 3         & 10 & 6 & 1 & 1 & 2 & 2    & 7 & 7 & 0 & 0 & 0 & 0      & 5 & 3 & 0 & 1 & 1 & 0 \\
    Not Helpful & 9 & 0 & 0 & 0 & 9 & 9         & 10 & 0 & 0 & 1 & 9 & 9    & 7 & 1 & 0 & 0 & 6 & 5      & 6 & 1 & 0 & 0 & 5 & 2 \\
    \bottomrule
    \addlinespace[1pt]
    \multicolumn{20}{@{}l@{}}{\makecell[l]{\footnotesize A feedback component can relate to multiple thoughts. For this reason, the sums of totals per action do not equal the sample sizes.}} \\
    \end{tabular}
    \end{center}
\end{table*}

\subsubsection{Additional Information Needed}

\textit{} \newline \textit{I don't understand:} In 8 cases, students stated they did not understand a piece of information, without specifying what information was missing. 

\textit{Syntax and Implementation:} Students were unsure about a syntactic implementation detail in 38 cases. They knew what to do, but not how to implement it. This was primarily observed with the \emph{Find Problems (KM)} (13.3\% of read \emph{KM} feedback components) and \emph{Strategic approach (SPS)} (16.7\% of read \emph{SPS} feedback components) feedback components (see \autoref{tab:thoughts}). 

\textit{Concepts and Explanations:} This category applies when a student requires an additional explanation or is uncertain about a programming concept (16 cases).
This occurred primarily with the \emph{Find Problems (KM)} component (9 cases) and the \emph{Next Step Hint (KH)} component (5 cases). For example, a student reads the \emph{KM} feedback, trying to understand an error. Yet, they still do not know how to fix them, and collect \emph{KH} feedback about the next step. However, neither of these components has provided a detailed explanation of, e.g., adding elements to a list. So the student thinks aloud, \say{What was append again?}, and consults the \emph{Explanations and Examples (KC)} component.

\textit{Strategy and next steps:} In 26 cases, students did not know what to do next. This was evident through their think-aloud process, which revealed that they lacked a clear strategy to move forward.
Notably, this occurred in 8 cases involving the \emph{Strategic approach (SPS)} feedback component (26.7\% of all read \emph{SPS} feedback components). This might be due to the \emph{SPS} component's positioning. It was typically the last one that students opened. Hence, there was no further feedback available. 
These situations were sometimes linked to frustration, leading students to consider abandoning the task: \say{The question is whether I should continue here. I just don't know what to do right now.} 

\subsubsection{Students Express Disagreement}

\textit{} \newline \textit{Disagree:} In 12 cases, students disagreed with or questioned the correctness of information provided by a tutoring feedback component for various reasons. For example, a student disagreed with a \emph{Next Step Hint (KH)} component because the student had already made changes to the code based on a previously read feedback component. In this case, a variable name was changed (from 'rates' to 'number'). 
The \emph{KH} feedback then referred to the original variable name (\say{In the if-statement, replace rates with the current grade, e.g., [Code Snippet]}). Then the student said: \say{But there is no 'rates' in the if-statement.} 

\textit{Irrelevant to the current problem:} In 25 cases, students stated that the information provided by the feedback component did not match the problem they were currently solving. This often happened when the feedback component focused on what students considered a trivial issue, such as a missing \texttt{return} statement, the use of \texttt{print} instead of \texttt{return}, or incorrect indentation. While the students did not directly neglect the feedback, they were focused on a different aspect of the problem-solving process.

\textit{Information already known:} Students thought aloud that they already knew the generated information (39 cases). In 24 instances, they stated that they knew it from a previous feedback component. In 15 instances, they did not specify the source of their knowledge. This category was observed most frequently with the \emph{Explanations and Examples (KC)} feedback components (18\% of 111 read \emph{KC} components). 

\subsubsection{Students' Explicit Judgments}
Overall, students judged the helpfulness of the feedback in their verbalized thoughts as \say{helpful} 26 times and \say{not helpful} 32 times.

\subsection{RQ 2: Relations of Processes in Feedback Engagement}

To investigate the relations between feedback collection, (visual) attention, (verbalized) thoughts, and (immediate) actions, we developed a coding scheme for student actions. For feedback components that were read by students, we coded subsequent actions (or inactions) using the categories \emph{Improvement (IMP)}, \emph{Setback (SET)}, \emph{Both (BOT)}, or \emph{No Action (NOA)} (see \autoref{tab:actions}). 
We also investigated whether the student collected an additional feedback component from the same GenAI-feedback request (\emph{COF}). We then analyzed the relations between the coded thoughts and the subsequent actions for each feedback component (see \autoref{tab:thoughts-feedback-action-outcome-relation}).
Overall, in 141 (82.0\%) of 172 GenAI feedbacks, at least one of the four components was collected, read, and resulted in an improvement (\emph{IMP}).

Students expressed understanding (185 verbalized thoughts) in conjunction with \emph{improvements (IMP)} in 76.2\% of instances (see \autoref{tab:thoughts-feedback-action-outcome-relation}).
An analysis of the coding frequencies relative to the total number of read components (see \autoref{tab:actions}) reveals that the \emph{Next Step Hint (KH)} component, compared to \emph{KM}, \emph{KC}, and \emph{SPS}, most frequently led to an \emph{Improvement} (in 52\% of the 73 read \emph{KH} instances). In 39.7\%, the \emph{KH} feedback component resulted in \emph{No Action (NOA)}. Students collected another feedback (\emph{COF}) component after reading \emph{KH} feedback in 31.5\% of cases.

\emph{Setbacks (SET)} were rare across all feedback components (see \autoref{tab:actions}).
Code changes that resulted in both an \emph{Improvement and a Setback (BOT)} were more frequent and sometimes occurred together with the \emph{Explanations and Examples (KC)} feedback component (12 cases, 10.8\% of all read \emph{KC}).
\emph{Setbacks (SET)} and mixed outcomes (\emph{BOT}) occurred more frequently with verbalized thoughts expressing understanding (43.6\% of 39 \emph{SET} and \emph{BOT} cases with verbalized thoughts) than with thoughts expressing disagreement (18.0\%), a need for additional information (25.6\%), or judgments of helpfulness (12.8\%).

There were 179 cases (47.1\% of the 380 read feedback components) of \emph{No Action (NOA)}. Thus, students acted upon the remaining 201 (52.9\%) of the 380 read feedback components, resulting in an \emph{Improvement} (160 cases), \emph{Setback} (13 cases), or both (\emph{Improvement and Setback}, 28 cases).

\emph{No Action} occurred in approximately 40\% of cases after reading \emph{Next Step Hint (KH)} components, and in roughly 50\% of cases for \emph{Find Problems (KM)}, \emph{Explanations and Examples (KC)}, and \emph{Strategic Approach (SPS)} components (relative to the total number of times each component was read, see \autoref{tab:actions}). 
Students often chose to \emph{Collect another Feedback Component} (212 cases). The frequency of this action, both in absolute terms and relative to how often each feedback component was read, decreases in the same order that the components were presented to the student in the user interface: From \emph{KM} (highest), to \emph{KC}, \emph{KH}, and last \emph{SPS} (lowest). This aligns with the observation that students generally followed the visual order of feedback components.

Students' verbalized thoughts and their subsequent actions were generally aligned. For example, when a thought belonged to the themes \emph{Additional information needed} or \emph{Students express disagreement}, it was often (69.5\%) followed by \emph{No Action}. In these situations, the most common subsequent step (79.2\%) was to \emph{Collect another Feedback Component}. This connection illustrates the internal validity of the data and methods.

\section{Discussion}
RQ1 addressed what students think when engaging with tutoring feedback.
With inductive coding of students' verbalized thoughts during feedback engagement, we identified four main themes: \emph{Students express understanding}, \emph{Additional information needed}, \emph{Students express disagreement},
\emph{Students' explicit judgments}.
Although students' (verbalized) thoughts indicated that they understood or agreed with most of the information provided, they also seemed to struggle somewhat during engagement.
For example, the category \emph{Additional information needed} was most prominent after reading the \emph{Find Problems (KM)} and \emph{Strategic Approach (SPS)} components, as students needed more information regarding \emph{Syntax and Implementation}.

By design, the feedback components (\emph{KM} and \emph{SPS}) did not provide concrete code examples. But even after reading a \emph{Next Step Hint (KH)} component with code snippets, some students required more specific guidance on exactly where to insert the suggested code into their solution.
Students thought about feedback as \emph{irrelevant to the current problem} when it focused on an aspect that they did not consider their primary obstacle. This reveals a limitation of feedback systems with predefined inputs, as students cannot steer the focus to their individual needs. We observed students developing strategies to avoid feedback that they predicted to be irrelevant to them. For example, students attempted to guide the feedback generation by providing pseudocode or comments.
Students sometimes thought aloud that they already knew the information presented by a feedback component (\emph{Information already known}). This occurred mostly for the \emph{Explanations and Examples (KC)} component, as it explained the same concepts relevant to a task repeatedly to students who requested multiple feedback components on a single task. This happened because the feedback generation process did not take into account a student's previous engagements.
Allowing for conversational follow-up questions could mitigate these issues by enabling students to direct the focus of the feedback or request more specific guidance tailored to their individual needs.

\begin{figure}[tb]
    \centering
    \includegraphics[width=1\linewidth]{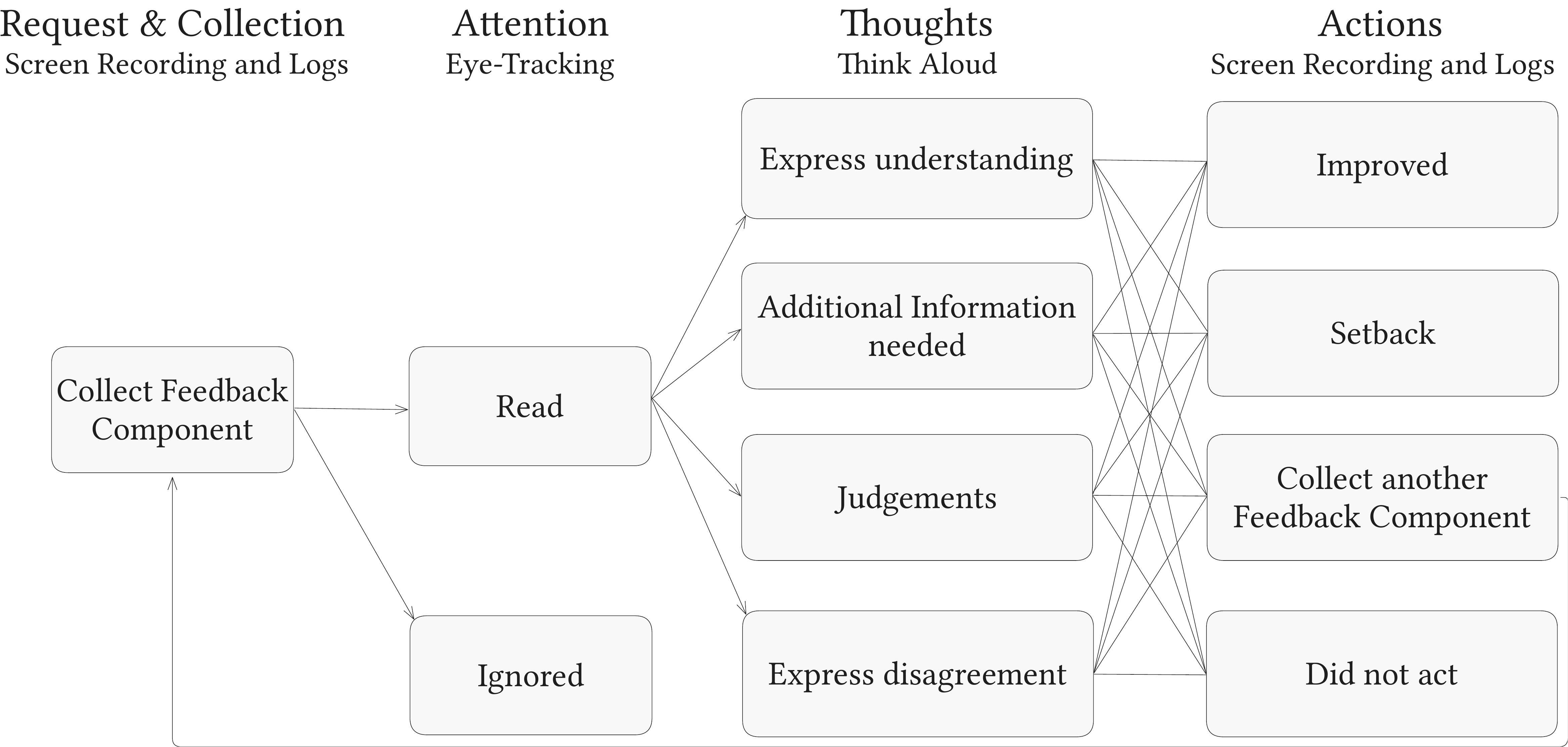}
    \caption{Feedback engagement processes and relations}
    \label{fig:Feedback-Engagement-Processes}
\end{figure}

For the second research question, we focused on students' actions related to their verbalized thoughts (see \autoref{fig:Feedback-Engagement-Processes}). 
Students demonstrated self-regulation \cite{butler.1995} in their active decisions on how they engage with tutoring feedback components. They only collected 55.8\% (384 of 688) of the generated feedback components -- not all of them. This may indicate that students are well aware of the type of feedback they need. However, students did not explicitly explain their choice of feedback components while thinking aloud. 
Once collected, almost all (380 of 384) feedback components got students' visual attention.
A student's decision not to act on feedback is not necessarily a sign of disengagement. They may have been deeply engaged in considering the feedback, but ultimately concluded that taking action was not adequate~\cite{price.2011}, as they are constantly evaluating external information (e.g., feedback) against their internal understanding. This is representative of the interplay of external and internal feedback loops \cite{narciss.2008}. The Think-Aloud Protocol enabled us to observe aspects of this (internal) evaluation.

If a student's internal evaluation of that information resulted in \emph{Additional information needed} or \emph{Students express disagreement}, they would often refrain from changing their code (\emph{NOA}). Instead, students continued with collecting another feedback component (\emph{COF}).
This successful internal evaluation prevented setbacks (actions coded as \emph{SET} and \emph{BOT}). 
In other cases, students' internal evaluation was not successful (from an external perspective). \emph{Setbacks} mostly co-occurred with verbalized thoughts coded as \emph{Agrees to Information} or \emph{Thoughts demonstrate understanding}. This may be due to a misunderstanding of the task, the feedback, or what is needed to solve the problem.
Mitigating this issue via technology or within a tool is a challenge. It, therefore, remains important to let students determine if they have understood feedback and perhaps foster their feedback literacy and self-regulation.

\section{Threats to Validity}
The limitations of the methodology should be noted when interpreting the results. These comprise, for example, the setting of the think-aloud study, which was artificial and supervised. Students may have behaved differently at home when not being observed~\cite{roethlisberger1939management}. 

In the questionnaire, we asked students about any negative or positive effects they experienced from wearing eye-tracking glasses or from the think-aloud protocol.
The eye-tracking equipment may have distracted some of the students (four students mentioned this limitation). Two others noted their vision was somewhat limited due to the glasses. Regarding the think-aloud protocol, six students noted that thinking aloud slowed down their thinking and concentration. At the same time, many others emphasized the more structured problem-solving while thinking aloud.

Moreover, this study is qualitative and thus explorative. Hence, the findings need to be confirmed for other student groups (e.g., in other countries) before they can be generalized, but this is common for the qualitative research paradigm. Yet, with a sample size of 20 students and 380 requested and read GenAI feedback components, we have reached saturation within the sample~\cite{boddy2016sample}.

Finally, feedback components often contained multiple pieces of information. Due to that, students sometimes uttered somewhat contradictory thoughts while processing a single feedback component. For example, a student might initially agree with one piece of information but later disagree with another piece of information from the same component (see \autoref{sec:results:thoughts}). 
It should also be noted that the fixed order of feedback components may have had confounding effects.

\section{Conclusions}
We evaluated how 20 introductory programming students engage with GenAI feedback via four different tutoring feedback components. Using Think-Aloud Protocols, Eye-tracking, and screen recordings, we investigated which tutoring feedback components were collected, got (visual) attention, and how students thought about and acted upon them.

Through qualitative analysis of students' verbalized thoughts during feedback engagement (RQ1), we identified four main themes. Students most frequently expressed understanding (45.5\% of verbalized thoughts) through agreement, verbalized thoughts demonstrating understanding, or restating the feedback in their own words. However, 21.6\% of verbalized thoughts indicated a need for additional information, e.g., regarding syntax implementation details and strategic guidance. Students also expressed disagreement (18.7\%) when feedback seemed irrelevant, was already known, or perceived as incorrect. The remaining 14.3\% of verbalized thoughts contained explicit judgments about the feedback's helpfulness.

By examining the relationship between students' verbalized thoughts and their immediate actions (RQ2), we found that students selectively collected only 55.8\% of the 688 generated feedback components. Almost all (99\%) of the collected components were read. Students acted upon 52.9\% of the feedback that they read.
When students' verbalized thoughts implying understanding, 76.2\% of related feedback components led to code improvements. Conversely, when thoughts expressed disagreement or a need for additional information, 69.5\% of related feedback components resulted in no immediate actions (code changes). Instead, students typically collected another feedback component (79.2\% of no-action cases). There was a high overall improvement rate: 82.0\% of GenAI feedback requests resulted in at least one improvement. Thus, students were able to collect and act upon AI-generated information through the self-directed selection of tutoring feedback components. 

This study contributes to our understanding of student engagement with tutoring feedback components by revealing some of the complex relationships within the feedback engagement process. Students' internal evaluation of external information was highlighted as a critical factor.
For educators and tool developers, it is therefore crucial to ensure that learners have understood the feedback they received. Allowing for conversational follow-up questions could, for example, help learners shift the focus of the feedback to their informational needs.
Future work should focus on developing pedagogical strategies to foster students' feedback literacy, enabling them to critically evaluate information and make informed decisions about their next steps.

\bibliographystyle{ACM-Reference-Format}
\balance
\bibliography{bib}
\end{document}